\documentstyle[12pt]{article} % the order of [12pt,ll] is important!

\def\Journal#1#2#3#4{{#1}{\bf #2}, #3 (#4)}

\def\PRS{\it Proc. Roy. Soc.~}
\def\ZP{\it Z. Phys.~}
\def\IJMP{Int. J. Mod. Phys.~}
\def\NPB{{\it Nucl. Phys.} {\bf B}}

\def\PRL{\it Phys. Rev. Lett.~}
\def\PRD{{\it Phys. Rev.} {\bf D}}

\def\adhoc{{\it ad hoc}}
\def\etal{{\it et al.}}
\def\ie{{\it i.e.}}

\begin{document}

\title{LOW ENERGY NUCLEON-NUCLEON SCATTERING\\ IN THE SKYRME MODEL}

\author{T. GISIGER, M.B. PARANJAPE}

\centerline{Groupe de physique des particules, D\'epartement de physique,}
\centerline{Universit\'e de Montr\'eal, C.P. 6128, 
succ. centre-ville, Montr\'eal,}
\centerline{Qu\'ebec, H3C 3J7, Canada}

\maketitle

\abstract{We present the study of the influence of the leading contribution
from the kinetic energy of the Skyrme Lagrangian on the scattering of low
energy nucleons. This classical kinetic energy for the 2-body system is
computed using the product ansatz and is correct for low energy, well separated
Skyrmions.
We quantize the rotational degrees of freedom of the Skyrmions in order to give
them the right spin and isospin values. We are then able to compute
analytically the scattering angles for some angular momenta polarizations.
}

\section{Introduction}

In this work\cite{ref:TGMBP2} we present some of the implications of
the Skyrme\cite{ref:Sk} model on the
scattering of two nucleons. We will first start with a short introduction to
the Skyrme model and its simplest solution, the Skyrmion. Then in the next
section we study the problem of the scattering of two Skyrmions using the
product ansatz and present the constraints imposed by this parametrization to
the scattering processes. The expression of the energy of the two Skyrmion
system is then written as an expansion in inverse powers of the separation
between the particles. We then briefly describe the approximation methods used
to find and solve the equations of motion of the system, namely the method of
Lagrange and the method of variation of constants. In order to study the
scattering of nucleons, we have to quantize the remaining degrees of freedom of
the system, namely the rotation of each Skyrmion so as to give them
the proper spin and isospin. This is done with the Bohr-Sommerfeld quantization
technique. We are then able to compute the scattering of two nucleons for
certain spin and isospin polarizations.

\section{The Skyrme model and the Skyrmion}

The Skyrme mode was introduced\cite{ref:Sk} in the 60's and is described by the
following Lagrangian:
\begin{equation}
{\cal L}_{sk} =
-{f_\pi^2 \over 4}\; tr(U^\dagger \partial_\mu U U^\dagger \partial^\mu U)
+ {1\over 32 e^2}\;tr( [U^\dagger \partial_\mu U,U^\dagger \partial_\nu U]^2)
\label{eq1}
\end{equation}
where $U(x^\mu)$ is an element of $SU(2)$ which represents a pseudoscalar
massless particle, namely the pion. $f_\pi$ and $e$ are phenomenological
parameters related to meson decay and low energy scattering, and
are theoretically computable using QCD.

The simplest stable solution of the Skyrme model is the Skyrmion:
\begin{equation}
U_s(\vec x) = e^{i F(r) \vec \tau\cdot\hat r}\label{eq2}
\end{equation}
where $\vec\tau=(\tau^1,\tau^2,\tau^3)$ are the three Pauli matrices and the
function
$F(r)$ is a \break monotonous function of $r$ which decreases from $\pi$ at the
origin, to 0 at infinity. $F(r)$ has to be computed numerically but falls off
at large distance like
$\kappa/r^2$. Since the Skyrmion field attains the $SU(2)$ identity when
$r\rightarrow\infty$, we can consider the three dimensional space $R^3$ to
be topologically compactified into a 3-sphere $S^3$. $U_s(\vec r)$ then defines
an application from an $S^3$ to the manifold $S^3$ of the $SU(2)$ group.
This type of
application is classified into disjoint homotopy classes labeled by the winding
number of the first 3-sphere onto the second, $N$, defined by
\begin{equation}
N = {1 \over 24 \pi^2} \int{d^3 \vec x\, \epsilon^{ijk} \;
tr(U^\dagger \partial_i U U^\dagger \partial_j U U^\dagger \partial_k U)}.
\label{eq3}
\end{equation}
An application belonging to one
class cannot be deformed continuously into another: it is
topologically stable. The Skyrmion has
winding number 1, rendering it stable against deformation into the vacuum
defined by $N=0$. It was Skyrme who first interpreted $N$ as the baryon
number.

The Skyrmion is a static solution. In order to set it into motion, we shift
its position by the time dependent vector $\vec R(t)$ and conjugate it by the
time dependent $SU(2)$ matrix $A(t)$:
\begin{equation}
U_1(\vec r,t)=A(t) U_s(\vec x-\vec R(t)) A^\dagger (t).\label{eq4}
\end{equation}
We then replace $U_1(\vec r,t)$ into the Skyrme Hamiltonian density and
integrate over all space. This gives the following expression for the energy:
\begin{equation}
E_{1} = M + {M\over 2} \dot{\vec R^2} + 2 \Lambda {\cal L}^a(A) {\cal
L}^a(A)\label{eq5}
\end{equation}
where we defined the generators of the right and left action
of the $SU(2)$ group:
\begin{equation}
{\cal R}^a(A)=-{i\over 2}\; tr[\tau^a A^\dagger \partial_0 A]\label{eq6}
\end{equation}
\begin{equation}
{\cal L}^a(A)=-{i\over 2}\; tr[\tau^a \partial_0 A A^\dagger].\label{eq7}
\end{equation}
$E_1$ has the familiar form of the energy of a translating and spinning
spherically symmetric rigid
body if $M$, of the order of 1 GeV, is interpreted as the mass, and $\Lambda$
as the moment of inertia.
The last term is the rotational energy, with ${\cal L}^a(A)
{\cal L}^a(A)$ equal to the square of the angular velocity. ${\cal L}^a(A)$ is
usually interpreted as the tensorial part of the isospin of the Skyrmion,
and ${\cal R}^a(A)$ as its spin.

It was shown by Atkins, Nappi and Witten\cite{ref:Witt}, that by quantizing the
rotational degrees of freedom $A(t)$ the characteristics of the nucleon could
be
reproduced
with an error varying between 10\% and 30\%. This guides us towards
investigating
the predictions of the Skyrme model in the baryon number 2 sector.

\section{The Skyrmion-Skyrmion interaction}

We will now attempt to compute the classical Lagrangian describing a pair of
Skyrmions in order to study their scattering. To
achieve this we need a parametrization of the pair of Skyrmions. The simplest
is the so-called ``product ansatz'' where the Skyrmion fields are multiplied
together:
\begin{equation}
\begin{array} {l}
U_2(\vec r,t)=\;U_1(Skyrmion\; 1) \; U_1(Skyrmion\; 2)
\\
\qquad\qquad=\;A(t) U_s(\vec x-\vec R_1(t)) A^\dagger (t)\;
B(t) U_s(\vec x-\vec R_2(t)) B^\dagger (t).
\end{array}\label{eq8}
\end{equation}
$A$ and $B$ are $SU(2)$ matrices representing the iso-orientation of the
Skyrmions and $\vec R_1$ and $\vec R_2$ their position. This parametrization
is not completely general and we have to
restrict the initial conditions in order to obtain physically correct results.

First, it is a well known result from numerical studies of the Skyrme
model\cite{ref:Wal} that Skyrmions deform when they come close to each other.
In fact, the
bound state of two Skyrmions, putatively the deuteron, has the form of a torus.
These deformations cannot be described by Eq. \ref{eq4}, let alone the product
ansatz, so we have to
consider only configurations where the distance $d=|\vec R_1(t)-\vec R_2(t)|$
is large.

Second, it was shown by N.S. Manton\cite{ref:Man} that only low energy soliton
systems can be described with a finite number of degrees of freedom. We refer
the interested reader to his article and will only say here that it is
physically evident that the number of degrees of freedom of a system
rises with
its energy. Then if we keep the energy of the system sufficiently low, the
system might excite only a finite number of modes. If we wish to describe our
system with the 12 degrees of freedom of the product ansatz (6 to define the
positions of the
particles, and another 6 for their iso-orientations), we have to consider
only low energy scattering.

We now give the expression of the energy for the pair of Skyrmions, in the
center of mass frame, as an expansion in inverse powers of the relative
distance $d$ which is assumed to be large at all times:
\begin{equation}
\begin{array} {l}
T = {1\over 4} M \dot{\vec d^{\;2}} + 2 \Lambda \bigl({\cal L}^a(A)\, {\cal
L}^a(A) + {\cal L}^a(B)\,{\cal L}^a(B)\bigr)
\\
\qquad\qquad\qquad+ {\Delta\over d} \epsilon^{iac}\epsilon^{jbd}\;{\cal R}^c(A)
\,{\cal R}^d(B)\;
\bigl(\delta^{ij}-\hat{d}^i \hat{d}^j\bigr)\, D_{ab}(A^\dagger B) +O(1/d^2).
\end{array}\label{eq9}
\end{equation}
$\Delta=2 \pi \kappa^2 f_\pi^2$ is a result of the integration over space,
$\hat d =\vec d/d$ and $D_{ab}(G)=1/2\;\hbox{tr}[\tau^a G \tau^b G^\dagger]$
is the
$3\times 3$ representation of the $SU(2)$ matrix $G$. This term, induced by the
kinetic energy of the Skyrme
Lagrangian, was found by us\cite{ref:TGMBP1} and independently by B.J.
Schroers\cite{ref:Schr}.

We note that this term has a structure similar to the spin-spin and tensorial
terms found in the traditional nuclear potentials. Also, the leading
contribution from the potential part of the Skyrme Lagrangian behaves as
$1/d^3$, thus is neglected to leading order of our expansion of the energy.
The kinetic energy defines a metric on the tangent space of the modulii space,
the motion implied by the Lagrangian simply follows the geodesics of this
metric. This is what we call the geodetic approximation.

\section{Equations of motion of the system and quantization}

We now have to find the equations of motion corresponding to the Lagrangian of
Eq. \ref{eq9}. Because of the very complicated form of the interaction term, we
have to resort to various approximation methods. We first choose the
observables
of the system, meaning, what we need to know from the scattering in order to
possibly compare the results with experimental data. We choose the relative
velocity
$\dot{\vec d}$, the tensorial part of the spins ${\cal R}^a(A)$ and
${\cal R}^a(B)$, and isospins ${\cal L}^a(A)$ and ${\cal L}^a(B)$.

The approximation method of Lagrange\cite{ref:Gold} is perfectly suited to
compute the time
derivatives of the observables without having first to derive the equations of
motion of $A$, $B$ and $\vec d$. The method uses the Poisson brackets formalism
and is based on the following principle. Let us consider the degree of freedom
$q^i$. Conjugate to $q^i$ is the momentum $p^i$ which can be written as an
expansion in powers of $1/d$:
\begin{equation}
p^i=p_0^i + \delta p^i(1/d).\label{eq10}
\end{equation}
Then the Poisson brackets, which are functions of the $q^i$ and the $p^i$,
also form an expansion in powers of $1/d$. If we denote by $C^k$ the
observables of the system, the time derivative of $C^k$ is
\begin{equation}
{d\over dt}C^k = \{C^k,H\} = \{C^k,H_0+H_I\}\label{eq11}
\end{equation}
where we have divided the Hamiltonian into a free $H_0$ (of order
$1/d^{\,0}$) and interaction $H_I$ part (of order $1/d$ and higher).
By dividing also the Poisson
bracket into a free and interaction part, we get
\begin{equation}
{d\over dt}C^k = \{C^k,H_0\}_I + \{C^k,H_I\}_0\label{eq12}
\end{equation}
plus higher order terms. By computing the various Poisson brackets involved, we
get the following set of 5 coupled equations:
\begin{equation}
\begin{array} {l}
{d\over dt}\dot d^k == -{2\Delta\over M d^2}\biggr[\delta^{ij}\hat d^k +
\delta^{jk}\hat
d^i +\delta^{ik}\hat d^j - 3\hat d^i \hat d^j \hat d^k\biggl] \epsilon^{iac}
\epsilon^{jbd} {\cal R}^c(A) {\cal R}^d(B) D_{ab}(A^\dagger B)
\\
{d\over dt}{\cal L}^k(A) =  {\Delta\over 2 M d} \epsilon^{iac} \epsilon^{jbd}
{\cal R}^c(A) {\cal R}^d(B) \bigl(\delta^{ij}-\hat d^i\hat d^j\bigr)
\epsilon^{kef} D_{fa}(A)D_{eb}(B) + \cdots
\\
{d\over dt}{\cal L}^k(B) =  {\Delta\over 2 M d} \epsilon^{iac} \epsilon^{jbd}
{\cal R}^c(A) {\cal R}^d(B) \bigl(\delta^{ij}-\hat d^i\hat d^j\bigr)
\epsilon^{kef} D_{ae}(A^\dagger )D_{fb}(B) + \cdots
\\
{d\over dt}{\cal R}^k(A) = - {\Delta\over 2 M d}\epsilon^{iac} \epsilon^{jbd}
{\cal R}^d(B)\bigl(\delta^{ij}-\hat d^i\hat d^j\bigr)
\\
\qquad\qquad\qquad\qquad\qquad\times\biggl[\epsilon^{kcf} {\cal R}^f(A)
D_{ab}(A^\dagger B) +
\epsilon^{kaf} D_{fb}(A^\dagger B) {\cal R}^c(A)\biggr] + \cdots
\\
{d\over dt}{\cal R}^k(B) = - {\Delta\over 2 M d}\epsilon^{iac} \epsilon^{jbd}
{\cal R}^c(A)\bigl(\delta^{ij}-\hat d^i\hat d^j\bigr)
\\
\qquad\qquad\qquad\qquad\qquad\times\biggl[\epsilon^{kdf} {\cal R}^f(B)
D_{ab}(A^\dagger B) +
\epsilon^{kbf} D_{af}(A^\dagger B) {\cal R}^d(B)\biggr] + \cdots
\end{array}\label{eq13}
\end{equation}
where the dots represent other complicated terms.
This is still too complicated to solve algebraically and we have to resort to
further approximations. We choose the approximation method of the
variation of constants\cite{ref:Gold}. Let us consider as an example the
following equation:
\begin{equation}
{d\over dt} \vec x(t) = f(\vec x(t),t)\label{14}
\end{equation}
and treat it not as an ordinary differential equation but more like an
iteration equation:
\begin{equation}
{d\over dt} \vec x(t) \simeq f(\vec x_0(t),t)\label{eq15}
\end{equation}
where $\vec x_0(t)$ is a trial function. Then if the trial function
$\vec x_0(t)$
is close to the true solution $\vec x(t)$ or if $f(\vec x_0(t),t)$ is small,
then a good estimation of the solution of the equation is:
\begin{equation}
\vec x(t) = \int^t_{-\infty} f(\vec x_0(t'),t') dt'.\label{16}
\end{equation}
In our case, $\vec x_0(t)$ is the free value of the $C^k$, so without the
interaction term. The approximation method should give accurate results if the
$C^k$ are slowly changing quantities. This is compatible with our working
hypothesis. This method decouples the 5 equations of motion and trivializes the
integration over time of the observables: all we have to do is replace in each
right hand side the free values for $\vec d$, $\dot{\vec d}$, ${\cal R}^a(A)$,
${\cal R}^a(B)$, ${\cal L}^a(A)$, ${\cal L}^a(B)$, $A$ and $B$, and integrate
over time each equation separately.

We will not consider here the time evolution of the spins and isospins. We will
only say that spin-flips and isospin-flips seem to occur, which are consistent
with spin 1 and isospin 1 particle exchange. We will only consider from now
on the equation for the relative momentum $\vec p = M/2 \dot{\vec d}$:
\begin{equation}
{d\over dt} p^i = -{\Delta\over d^2}\biggr[\delta^{ij}\hat d^k +
\delta^{jk}\hat
d^i +\delta^{ik}\hat d^j - 3\hat d^i \hat d^j \hat d^k\biggl] \epsilon^{iac}
\epsilon^{jbd} {\cal R}^c(A) {\cal R}^d(B) D_{ab}(A^\dagger B)\label{eq17}
\end{equation}
in order to extract the scattering angles.

To summarize, we now only have the equation for ${\vec p}$ to solve. This is
done by replacing in the right hand side of the equation the free values
of $\vec d$ and the matrices $A$ and $B$ which define also the spin and
isospin of the Skyrmions, and to integrate over time.

So far we have only talked about classical scattering. If we want to study the
physics of nucleons, we have to quantize the remaining degrees of freedom,
\ie$\;$the rotation of
each Skyrmion. This means finding the matrices $A$ and $B$ which give the
Skyrmions spin and isospin $1/2$. The Bohr-Sommerfeld method is perfectly
suited to this task. In the well known problem of the hydrogen atom, the sum of
the action variables
\begin{equation}
J_i = \oint p_i dq_i\label{eq18}
\end{equation}
is set equal to a multiple of the Plank constant $h$. This fixes the allowed
values for the angular and linear velocities, thereby quantizing the energy
and angular momenta of the states. In our case, the same program is applied
to the time derivatives of the angles with which the matrices $A$ and $B$ are
expressed. Here are the main steps of this procedure.

Let us consider the Lagrangian for a single Skyrmion and express the matrix
$A$ of its iso-orientation as a function of 3 Euler angles $\alpha$, $\beta$
and $\gamma$:
\begin{equation}
A=a_0 + i \vec a\cdot\vec\tau=e^{-i\alpha \tau_3/2}\,e^{-i\beta
\tau_2/2}\,e^{-i\gamma \tau_3/2}\label{eq19}
\end{equation}
giving
\begin{equation}
L = - M + {1\over 2}\Lambda\,\bigl[ \dot \alpha^2+ \dot \beta^2 +\dot
\gamma^2+ 2 \dot \alpha \dot \gamma \cos \beta\bigr].\label{eq20}
\end{equation}
The equations of motion are satisfied if $\alpha$ and $\gamma$ are linear
functions of time $t$ and if $\beta$ is equal to 0 or $\pi$.
The action variables for the angles $\alpha$ and $\gamma$ are then readily
computed and give:
\begin{equation}
\begin{array} {l}
J_\alpha\equiv\oint p_\alpha d\alpha =2 \pi\Lambda\bigl[
\dot \alpha
+ \dot \gamma \cos \beta\bigr]=-4\pi\Lambda{\cal L}^3(A)\equiv -2\pi I_3
\\
J_\gamma\equiv\oint p_\gamma d\gamma  =2\pi\Lambda\bigl[ \dot \gamma + \dot
\alpha \cos \beta\bigr]=-4\pi{\cal R}^3(A)\equiv 2\pi S_3
\end{array}\label{eq21}
\end{equation}
using Eq. \ref{eq6} and Eq. \ref{eq7}, and following the convention used by
Adkins\cite{ref:Witt} \etal$\;$
for the component of the spin and isospin along the $z$ axis
\begin{equation}
\begin{array} {l}
S_3 = 2 \Lambda {\cal R}^3(A)
\\
I_3 = -2 \Lambda {\cal L}^3(A)
\end{array}\label{eq22}
\end{equation}
respectively. The Bohr-Sommerfeld quantization condition states that the sum
$J_\alpha+J_\gamma$ is equal to an integer multiple of $h$. This fixes the
value of the angular speeds $\dot\alpha$ and $\dot\gamma$ so as to give, in
our particular case, spin and isospin $\pm 1/2$. The expression for the
matrix $A$
giving the proper quantum numbers is then obtained by extracting the values of
$\alpha(t)$ and $\gamma(t)$ from Eq. \ref{eq21}, after choosing the value
of $\beta$ which is either 0 or $\pi$.
\noindent
If $\beta=0$, solving for the angles $\alpha$ and $\beta$ gives the matrix
\begin{equation}
A = e^{i (\omega t + \phi_0) \tau^3/2}.\label{eq23}
\end{equation}
$\omega$ is the quantized quantity here, and has an absolute value of roughly
100 MeV for the nucleon. This matrix
represents the state $|p\downarrow>$ if $\omega>0$ and $|n\uparrow>$ if
$\omega<0$.
\noindent
If $\beta=\pi$, then we obtain
\begin{equation}
A = -i e^{-i (\omega t + \psi_0)\tau^3/2}\,\tau^2\,
e^{i (\omega t + \psi_0)\tau^3/2}\label{eq24}
\end{equation}
which represents the state $|p\uparrow>$ if $\omega>0$ and $|n\downarrow>$
if $\omega<0$.

Before going further we note that the time dependence of the matrices $A$ when
$\beta=0$ or $\beta=\pi$ are very different. This will have important
consequences on the nucleon-nucleon scattering.

\section{Nucleon-Nucleon Scattering}

We now have all the necessary tools to solve our problem. To solve Eq.
\ref{eq17} for a
particular scattering process, we only replace the corresponding matrices $A$
and $B$ in the right hand side of the equation, as well as the free value of
$\vec d$:
\begin{equation}
\vec d = (v t,\gamma,0)\label{eq25}
\end{equation}
which describes a relative particle traveling along the $x$ axis but with an
impact parameter $\gamma$ along the $y$ axis. To respect the restrictions
underlined in section 3, $v$ is chosen small to guaranty low energy, and
$\gamma$ large, so as to keep the particles far from each other.
In what follows, we
choose $z$ as the axis of polarization of the angular momenta. This gives
2-dimensional trajectories. Other polarization axes give complicated
3-dimensional motion best studied numerically.

Since in our approximation scheme the spins and isospins are constant, the only
time dependence on the right hand side of Eq. \ref{eq17} comes from $d$, $\hat
d$ and $D_{ab}(A^\dagger B)$. It is then predictible that
scattering processes separate into two cases, depending on
whether or not $A^\dagger B$ is time independent.
\begin{center}
{\bf $D_{ab}(A^\dagger B)$ time independent:}
\end{center}
\begin{eqnarray}
&
\begin{array} {l}
p\uparrow p\uparrow
\\
n\downarrow n\downarrow
\\
p\downarrow p\downarrow
\\
n\uparrow n\uparrow
\end{array}
&
{d\over dt} p^k = - {\Delta\omega^2\over d^2} \cos({2\delta}) \hat d^k
\\
&
p\uparrow p\downarrow
&
\begin{array}{l}
{d\over dt} p^k = - {\Delta\omega^2\over d^2}\bigl[
\hat d^k + 4 r^k \hat r\cdot\hat d - 6 \hat d^k (\hat r\cdot\hat d)^2 \bigr]
\\
\qquad\qquad\qquad\qquad\hat r^k = (-\sin(\delta),\cos(\delta),0)
\end{array}
\\
&
n\uparrow n\downarrow
&
\begin{array}{l}
{d\over dt} p^k = - {\Delta\omega^2\over d^2}\bigl[
\hat d^k + 4 r^k \hat r\cdot\hat d - 6 \hat d^k (\hat r\cdot\hat d)^2 \bigr]
\\
\qquad\qquad\qquad\qquad\hat r^k = (\sin(\delta),-\cos(\delta),0)
\end{array}
\end{eqnarray}
where $\delta=\phi_0^A-\phi_)^B$. The scattering angle defined as the angle
between the initial $(t=-\infty)$ and final $(t=+\infty)$ momenta is the same
for all these different processes and readily computed:
\begin{equation}
\begin{array}{l}
\cos\theta = {\vec p(+\infty)\cdot\vec p(-\infty)\over
                         |\vec p(+\infty)||\vec p(-\infty)|}
\\
\quad\quad={M\gamma v^2\over 4 \Delta}
{1\over\bigl( {M^2\gamma^2 v^4\over 16
\Delta^2} + \omega^4 \cos^2 2\delta \bigr)^{1/2}}.
\end{array}\label{eq29}
\end{equation}
\begin{center}
{\bf $D_{ab}(A^\dagger B)$ time dependent:}
\end{center}
\begin{eqnarray}
&
\begin{array} {l}
p\uparrow n\downarrow
\\
p\downarrow n\uparrow
\end{array}
&
{d\over dt} p^k = {\Delta\omega^2\over d^2} \cos(4\omega
t+2\epsilon)\hat
d^k
\\
&
p\uparrow n\uparrow
&
\begin{array}{l}
{d\over dt} p^k = {\Delta\omega^2\over d^2}\bigl[
\hat d^k + 4 r^k \hat r\cdot\hat d - 6 \hat d^k (\hat r\cdot\hat d)^2 \bigr]
\\
\qquad\qquad\qquad\qquad\hat r^k = (-\sin(2\omega t + \epsilon),
\cos(2\omega t + \epsilon),0)
\end{array}
\\
&
p\downarrow n\downarrow
&
\begin{array}{l}
{d\over dt} p^k = {\Delta\omega^2\over d^2}\bigl[
\hat d^k + 4 r^k \hat r\cdot\hat d - 6 \hat d^k (\hat r\cdot\hat d)^2 \bigr]
\\
\qquad\qquad\qquad\qquad\hat r^k = (-\sin(2\omega t + \epsilon),
-\cos(2\omega t + \epsilon),0)
\end{array}
\end{eqnarray}
where $\epsilon=\phi_0^A+\phi_0^B$. For those scattering states, the right hand
side of the equations oscillates very quickly and the scattering angle is
suppressed by the very small factor
\begin{equation}
\sim e^{-({\omega\gamma\over v})}\label{eq33}
\end{equation}
since $\omega$ and $\gamma$ are large while $v$ is small.

We underline that this is the first analytical
calculation of nucleon-nucleon scattering from
essentially first principles, without recourse to \adhoc$\;$  models or
potentials. We emphasize that the Skyrme model is in principle derivable from
QCD and $f_\pi$ and $e$ are, as such, calculable parameters and, in that sense
this is also a QCD calculation. To calculate the classical scattering
cross-section we need to compute the scattering for all different polarizations
relative to the initial scattering plane. This would comprise a different
project which would probably be best achieved by numerical methods. Therefore
we are unable at this point to make a direct comparison with the experiment.

A peculiar aspect of our treatment is the strong dependence of the scattering
angle on the variables $\delta$ and $\epsilon$ which are just
the phase lag between the rotation of the Skyrmions at some time $t$ before the
scattering. These act as hidden variables, and have to be measured after the
scattering or averaged over.

\section*{Acknowledgements}

This work supported in part by NSERC of Canada and FCAR of Qu\'ebec.

\end{document}